\documentclass[english]{article}
\usepackage[T1]{fontenc}
\usepackage[latin9]{inputenc}
\usepackage{stmaryrd}
\usepackage{babel}
\begin{document}
\title{Marginal probabilities in boson samplers with arbitrary input states}
\author{Jelmer Renema}
\maketitle
\begin{abstract}
With the recent claim of a quantum advantage demonstration in photonics
by Zhong \emph{et al}, the question of the computation of lower-order
approximations of boson sampling with arbitrary quantum states at arbitrary
distinguishability has come to the fore. In this work, we present
results in this direction, building on the results of Clifford and
Clifford. In particular, we show:

1) How to compute marginal detection probabilities (i.e. probabilities
of the detection of some but not all photons) for arbitrary quantum
states.

2) Using the first result, how to generalize the sampling algorithm
of Clifford and Clifford to arbitrary photon distinguishabilities
and arbitrary input quantum states.

3) How to incorporate truncations of the quantum interference into
a sampling algorithm.

4) A remark considering maximum likelihood verification of the recent
photonic quantum advantage experiment.
\end{abstract}

\section{Introduction}

In a recent landmark result, Zhong \emph{et al} \cite{Zhong2020}
claimed the first demonstration of a quantum advantage in photonics.
This result immediately led to discussion \cite{Kalaicomment} focussing
on the extent to which the samples which this experiment produced
truly manifest large-scale quantum interference, and whether the sampling
carried out by the photonic device could be efficiently classicaly
simulated. 

This discussion focussed on \emph{marginal probabilities}, i.e. the
probability to observe some $k\leq n$ photons in a particular output
configuration $\phi_{k},$ summed over all possible outcomes for the
remaining photons. Informally, these probability distributions answer
the question `what is the probability to see a photon \emph{here,
here }and \emph{here}, irrespective of where the other photons go?' 

The reason for this interest is that unlike in random circuit sampling
\cite{Boixo2017}, marginal distributions carry some information about
the overall large-scale interference process, and can therefore be
used as a diagnostic tool, to build up evidence of the success of
a quantum advantage experiment. Specifically, $k$-th order marginals
(i.e. marginal distributions involving $k$ particles) carry information
about the interference processes of $k$ and fewer photons. 

Applying these ideas, the authors of \cite{Zhong2020} showed that
their experiment has first and second order marginal distributions
roughly consistent with the theoretical predictions. Furthermore,
the authors generated samples which have the first order marginals
correct (i.e. in agreement with the noiseless case) but not the higher
order marginals. The authors of \cite{Zhong2020} introduced a statistical
test (which has since received the name `CHOG') which essentially
tests which of two sets of samples is most likely to come from to
the ideal distribution. They showed that the first-order marginal
samples lose this test against the experimental samples. Kalai suggested
\cite{Kalaicomment}, based on an earlier paper of his and Kindler,
\cite{Kalai2014} that a series of samples which in addition to the
first-order marginals have up to $k$ order marginals correct would
win against the experimental samples. There was a subsequent counterclaim
\cite{Chaoyangcomment} by the authors of \cite{Zhong2020} that for
the case of $k=2,$ these samples also fail to beat the experimental
samples at the CHOG test. The issue of what would happen when comparing
higher marginals to the experiment is currently stuck at the question
of how to generate samples which have the correct third and higher
order marginal distributions. 

In this work, we resolve this issue by showing how to generate samples
which have the correct $k$-th order marginals for any $k$, for an
arbitrary quantum state, with arbitrary properties of the photons
in that state. This discussion builds on a rich literature of how
to approximate boson sampling with imperfections \cite{Kalai2014,Aaronsonlostphotons,Arkhipov15,Moylett2019,Oszmaniec2018,Patron2017,Qi2019,RahimiKeshari2016,Renema2018a,Renema2018b,Renema2020GBS,Shchesnovich2014,Shchesnovich2015a,Shchesnovich2015b,Shchesnovich2017,Shchesnovich2019a,Shchesnovich2019b,Shchesnovich2020,vdMeer2020}.

\section{The setting}

Consider a linear optical system of $N$ modes described by a transmission
matrix $U$. At the input of this optical system, an arbitrary quantum
state of light $|\Psi\rangle$ is impingent\footnote{$|\Psi\rangle$ may be entangled, and need not be a state of definite
overall photon number.}. Furthermore, we assume that there is photodetection in the Fock
basis, i.e. at the detector the set of measurement operators $\Pi_{i}=|i\rangle\langle i|$
is applied to each output mode of the interferometer. 

We wish to accomplish the task of computing marginal probabilities
for arbitrary input quantum states $|\Psi\rangle$ and arbitrary mutual
distinguishabilities (wavefunction overlaps) of the input photons,
and ultimately generating samples which follow these distributions.
To simplify notation, we will assume that the photons in this quantum
state have equal mutual wave function overlaps i.e. $x_{ij}=\langle\psi_{i}|\psi_{j}\rangle=x+(1-x)\delta_{ij}$,
but our results can be trivially generalized to the case of unequal
wavefunction overlaps\footnote{In this work, we will consider mutual distinguishability as the main
source of imperfections. The relation between this noise source and
noise on the interferometer as treated in \cite{Kalai2014} is discussed
by Shchesnovich in \cite{Shchesnovich2019a}}. 

As a special case, for $|\Psi\rangle=|1\rangle^{\varotimes n}|0\rangle^{\varotimes N-n},$
this is the boson sampling problem as proposed by Aaronson and Arkhipov
\cite{Aaronson2011}. In that case, the probability of observing a
given outcome $\phi$ is given by $P(\phi)=\mathrm{|Perm(M_{\xi,\phi})|^{2},}$where
$M_{\xi,\phi}$ is the submatrix of $U$ connecting the first $n$
input modes of $U$ to the set of output modes $\phi$ of interest,
and $\mathrm{Perm}$ is the permanent function\footnote{We may renumber the input without loss of generality so that the first
$n$ modes contain a photon}. $\xi=(1,2,...n)$ is the vector of input modes containing a photon,
which is introduced for consistency with later notation. For this
case, Clifford \& Clifford \cite{Clifford2017} solved the question
of how to compute marginal probabilities, which they used to construct
a sampling algorithm which generates samples at close to the computational
cost of evaluating a single permanent. We will import their results
to the broader setting of arbitrary distinguishabilities and input
quantum states.

\section{Preliminaries}

We follow the approach of \cite{Renema2020GBS} and project the input
state $|\Psi\rangle$ onto the subspace $\Xi$ of the Fock basis at
the input of the interferometer with exactly $n$ incident photons,
i.e. for all |$\xi\rangle\in\Xi,$$\langle\xi|\hat{N}|\xi\rangle=n$,
where $\hat{N}=\sum\hat{n}$ is the multimode photon number operator\footnote{See \cite{Renema2020GBS} for a more formal derivation}. 

In the rest of this work, we will assume the set $\{|\xi\rangle\}$
forms an orthonormal basis of $\Xi$, and that they are chosen such
that each $|\xi\rangle$ is a product state over the input modes ,
i.e. $|\xi\rangle=\prod_{i=1}^{N}|m_{i}\rangle,$where $|m_{i}\rangle$
is a Fock state in mode $i$ with $m$ photons, and the condition
that $|\xi\rangle$ lies in $\Xi$ enforces $\sum_{i=0}^{N}m_{i}=n$.
Note that this can always be done, since different product states
of this form are automatically orthonormal, and a simple counting
argument guarantees they form a basis. Furthermore, we will use the
notation that $\xi$ is the vector of Fock-state inputs corresponding
to $|\xi\rangle.$ We will use $\Xi$ both for the subspace and for
the set of all valid assignments of $\xi$ corresponding to that subspace.

Note that since the photon number is conserved by the interferometer,
conditional on observing $n$ photons and barring photon loss\footnote{In the presence of photon loss, the probability is given by a classical
sum over various $\Xi_{j}$ for $j>n$, so \emph{mutatis mutandis
}the discussion below holds in the presence of photon loss as well.}, the dynamics of an $n$-photon detection event are entirely governed
by interference in the subspace $\Xi.$ The probability to observe
some outcome $\phi$ is then given by \cite{Renema2020GBS}:

\begin{equation}
P(\phi)=\sum_{\xi\in\Xi}\sum_{\chi\in\Xi}c_{\xi}c_{\chi}^{\dagger}\mathrm{Perm}(M_{\xi,\phi})\mathrm{Perm(}M_{\chi,\phi})^{\dagger},
\end{equation}

where $c_{\xi}=\langle\Psi|\xi\rangle/\sqrt{\mu(\xi)\mu(\phi)}$,
$\mu$ is the multiplicity function, $\mathrm{Perm}$ is the permanent
function, and $M_{\xi,\phi}$ is the submatrix that connects the modes
in $\xi$ which contain a nonzero number of photons with the output
modes of interest. In what follows, for simplicity and following the
constraints of the experiment \cite{Zhong2020}, we will assume $\mu(\phi)=1,$
i.e. all $n$ photons emerge from distinct modes.

As an example of how to construct the set of $\xi,$ consider Gaussian
boson sampling (GBS). In that case, the input state is given by a
product state of pairs of modes over a given subset modes, with the
remaining ones empty, i.e. $|\Psi\rangle=|\psi\rangle^{\varotimes n/2}|0\rangle^{N-n},$
with $|\psi\rangle=\frac{1}{\cosh(r)}\sum_{n=0}^{\infty}(\exp(-i\phi)\tanh(r))^{n}|n,n\rangle$
the \emph{two-mode squeezed state, }where $r$ and $\phi$ form the
\emph{squeezing parameter} via $\zeta=re^{i\phi}.$ As an example,
for a three-squeezer boson sampling experiment detecting four photons,
the complete set of $\xi$ would be $(1,2,1,2)$, $(1,2,3,4)$, $(1,2,5,6)$,
$(3,4,3,4),$$(3,4,5,6)$, $(5,6,5,6),$ i.e. all $6$ ways of chosing
2 pairs of 3 possible sources with repetition (i.e. ${3+2-1 \choose 2}$
ways). Interference between these different emission processes has
been observed in experiments. In the case of GBS, for equal squeezing
parameters on all sources, these $\xi$ are all equiprobable, which
follows from the exponential distribution of pairs.

\subsection{Results from Clifford \& Clifford}

We introduce a few results and techniques from \cite{Clifford2017}. 

First, to compute the marginal output distributions, we switch to
the expanded sample space notation. In this formalism, we remove the
notational restriction that the vector specifying the location of
the output photons at $\phi$ needs to be ordered. Normally, if we
give a list of the output modes of a given output configuration $\phi$
of a boson sampling experiment, we require that that list is ordered,
i.e. that $\phi(i)\leq\phi(i+1)$. Removing this restriction lifts
the spurious correlation between elements of $\phi$ arising from
this ordering. In the expanded sample space formalism, we have that
\cite{Clifford2017}: 
\begin{equation}
P(\phi_{u})=\frac{\mu(\phi)}{n!}P(\phi),
\end{equation}
where $\phi_{u}$ stands for an unordered version of $\phi,$

Next, we introduce two key expressions from \cite{Clifford2017} (rephrased
in our notation). First:
\begin{equation}
\mathrm{Perm(M_{\xi,\phi})\mathrm{Perm}(M_{\chi,\phi}^{\dagger})=\sum_{\sigma\in S}\mathrm{Perm(M_{\xi,\phi}\circ M_{\sigma(\chi),\phi}^{\dagger})}},
\end{equation}
Where the sum runs over all permutations and $\circ$ is an elementwise
product. Secondly:

\begin{equation}
\sum_{\phi\in\Phi}\mathrm{Perm}(M_{a,\phi}\circ M_{b,\phi}^{\dagger})=m!\delta(a,b),
\end{equation}
where $m$ is the size of $M$, and $\Phi$ is the set of all valid
assignment of the output modes (i.e. all possible outcomes), and $\delta$
is an elementwise Kronecker delta function. This result follows from
the orthonormality of the rows and columns of $U.$ Note that combining
equations 1-4 guarantees that $P(\phi)$ is normalized for any choice
of $|\Psi\rangle.$ 

\section{Results}

We obtain marginal distributions by applying equations 2-4 to 1. First,
we apply eqn 2 and 3. 

\begin{eqnarray}
P(\phi_{u}) & = & \frac{1}{n!}\sum_{\xi\in\Xi}\sum_{\chi\in\Xi}c_{\xi}c_{\chi}^{\dagger}\sum_{\sigma}\mathrm{Perm}(M_{\xi,\phi}\circ M_{\sigma(\chi),\phi}),
\end{eqnarray}
then, we apply Laplace expansion along the output modes, to split
up the permanent into those photons over which we will marginalize
and those which will remain. Note that without loss of generality,
we can pick the unmarginalized photons to be the first $k$ ones.
This results in:

\begin{equation}
P(\phi_{u})=\frac{1}{n!}\sum_{\xi\in\Xi}\sum_{\chi\in\Xi}c_{\xi}c_{\chi}^{\dagger}\sum_{\sigma}\sum_{\rho}\mathrm{Perm}(M_{\rho(\xi),\phi_{1:k}}\circ M_{\rho(\sigma(\chi)),\phi_{1:k}})\mathrm{Perm(M_{\rho(\xi),\phi_{k+1:n}}\circ M_{\rho(\sigma(\chi)),\phi_{k+1:n}})},
\end{equation}

where $\rho$ is the set of all $k$ combinations out of $n$, and
$\bar{\rho}$ is the complement of $\rho$.

Next, we apply eqn 4, summing over the $k+1$st to $n$th photon.
Using the independence of the various sums, this gives:
\begin{equation}
P(\phi_{k})=\frac{k!}{n!}\sum_{\xi\in\Xi}\sum_{\chi\in\Xi}c_{\xi}c_{\chi}^{\dagger}\sum_{\sigma}\sum_{\rho}\mathrm{Perm}(M_{\rho(\xi),\phi_{1:k}}\circ M_{\rho(\sigma(\chi)),\phi_{1:k}})\mathrm{\delta(\bar{\rho}(\xi),\bar{\rho}(\sigma(\xi)))}.
\end{equation}

We will now introduce partial distinguishability to the problem. Interestingly,
it is straightforward to introduce partial distinguishability to equation
7. The reason for this is that in the model of partial distinguishability
outlined in Section 1, each term that doesn't correspond to a fixed
point picks up a factor $x,$ since it is sensitive to the wavefunction
overlap between two different photons (fixed points pick up a factor
1 by construction). Since the delta function in eqn 7 tests whether
all marginalized photons correspond to fixed points of the partial
permutation ${\xi \choose \sigma(\rho)}$, this means that all non-fixed
points must be in the first $1...k$ modes. Therefore, the expression
at partial distinguishability reads: 

\begin{eqnarray}
P(\phi) & = & \frac{k!}{n!}\sum_{\xi\in\Xi}\sum_{\chi\in\Xi}c_{\xi}c_{\chi}^{\dagger}\sum_{j=0}^{k}x^{j}\sum_{\sigma_{j}}\sum_{\rho}\mathrm{Perm}(M_{\rho(\xi),\phi_{1:k}}\circ M_{\rho(\sigma(\chi)),\phi_{1:k}})\mathrm{\delta(\bar{\rho}(\xi),\bar{\rho}(\sigma(\xi)))},\nonumber \\
\end{eqnarray}
Which is the central result of this work. In this expression, $\sigma_{j}$
is a permutation with $k-j$ fixed points. 

Equation 8 shows the interplay between marginalizing over photons
and partial distinguishability. It shows that there are only two possibilities:
either all unfixed points lie $\rho$, in which case the effect of
partial distinguishability doesn't change ($j$-th order interference
remains at $j$-th order) or there are one or more unfixed points
in $\bar{\rho},$ in which case the entire term sums to zero. This
proves the claim in the introduction that marginal distributions can
be used to test lower-order interference up to $k$-th order. This
was noted previously by Walschaers \emph{et al} for two-point correlators,
which are closely related to second order marginal probabilities \cite{Walschaers2016}. 

\section{From computing output probabillities to sampling}

Eqn 8 only allows us to compute marginal probabilities, however what
we wish to do is sample from the distribution given by eqn 1, which
has these marginal probabilities, or even sample from some approximation
of that distribution. If we wish to generate samples from eqn 1, without
any further approximations, things are fairly straightforward. Eqn
8 allows us to use the procedure of Clifford and Clifford, which relies
on the repeated application of the chain rule for probabilities to
place the photons sequentially. We place the first photon according
to $P(\phi_{1})$, which results in placement of that photon in the
$j-$th mode. The second photon is placed according to $P(\phi_{2})|_{k_{1}=j}$,
i,e. $k=2$ with the first photon fixed in the position where we placed
it in the previous step, and so on. We repeat this until we have placed
$n$ photons. If we are sampling over a state $|\Psi\rangle$ of indeterminate
photon number, we can place an outer loop around this procedure in
which we first draw $n$ according to the appropriate distribution
given by $|\Psi\rangle,$ and then proceed as before. 

If we wish to approximate eqn 8 as only low-order interference, we
can do this by truncating the sum over $j$ to some value $j_{\mathrm{max}}.$
From previous work \cite{Kalai2014,Renema2018a}, we know that the
distribution given by these probabilities is close in $L_{1}$ distance
to the actual distribution, if the experiment has strong enough imperfections
(i.e. low mutual indistinguishability or high photon loss). In that
case, the $j$-th order marginal approximation $P'_{j_{\mathrm{max}}}(\phi_{k})$
is given by:
\begin{eqnarray}
P_{j_{\mathrm{max}}}'(\phi) & = & \frac{k!}{n!}\sum_{\xi\in\Xi}\sum_{\chi\in\Xi}c_{\xi}c_{\chi}^{\dagger}\sum_{j=0}^{j_{\mathrm{max}}}x^{j}\sum_{\sigma_{j}}\sum_{\rho}\mathrm{Perm}(M_{\rho(\xi),\phi_{1:k}}\circ M_{\rho(\sigma(\chi)),\phi_{1:k}})\mathrm{\delta(\bar{\rho}(\xi),\bar{\rho}(\sigma(\xi)))}.\nonumber \\
\end{eqnarray}

Since there are $n^{j}$ permanents with $j$ unfixed points, the
inner sum now sums over polynomially many terms.

By incorporating this approximation into the Clifford algorithm, we
have introduced some additional error. The reason for this is that
since equation 9 does not correspond to a physically realizible Gram
matrix for the wave function overlap of all the photons, it is no
longer guaranteed that the approximate probabilities will be positive.
The solution to this is to set all negative probabilities to be zero.
However, since the unmodified distribution is normalized, this influences
the normalization. We leave the analysis of this effect to future
work. 

\section{Closing remarks }

We conclude with a few points: 

\subsection{Application of Gaussian boson sampling, first and second order marginals}

As an illustration, we carry out some of the work of applying this
result to Gaussian boson sampling. For cases with a high degree of
symmetry, it is not necessary to explicitly enumerate all $\xi\in\Xi$
and $\chi\in\Xi$. For example, for Gaussian boson sampling, we can
use the pairwise product structure of the modes to simplify our computation.
Furthermore, we will assume equal squeezing in all modes, leading
to $c_{\xi}=c_{\chi}$ for all $\xi$ and $\chi$.

For the first order marginal, the pairwise structure of the state
imposes that $\xi=\chi,$ $\sigma=I$, where $I$ is the identity
permutation, since if $\xi\neq\chi$ they must differ in at least
two places. Hence by symmetry we have
\[
P(\phi_{1})=\frac{1}{m}\sum_{j=1}^{m}|U_{j,k}|^{2},
\]

where $m$ is the number of modes connected to a squeezer and $k$
enumerates the output modes.

For the second order marginals, we have two kinds of terms: those
where $\xi=\chi,$ and those where $\xi$ and $\chi$ differ by one
pair. The first case gives rise to terms analogous to boson sampling
with Fock states. The second case gives rise to phase-dependent terms,
which have no analog in regular boson sampling. The presence of these
terms was noted as a special feature of Gaussian boson sampling previously,
by Phillips \emph{et al }\cite{Phillips2019}\emph{. }We now see how
these terms arise: they correspond to quantum interference of different
histories of how the photons were emitted. Higher order marginals
will have the corresponding higher order intereference terms, i.e.
at the fourth marginal $\xi$ and $\chi$ can differ in two pairs,
and so on. 

\subsection{Maximum likelihood estimation}

In \cite{Renema2020ML}, the idea was floated to use maximum likelihood
estimation to test the quality of a boson sampler. The present results
show that this can also be done via the marginal distributions, since
they contain information about the mutual distinguishability. How
to interpret such an analysis would depend on the level of security
assumptions which we are willing to make: if we consider these measurements
a trusted characterization experiment, then the information about
all $k>2$ photon interference processes is contained in the second-order
marginals. 

However, with the procedure outlined above, it would be possible to
generate outcomes which have the higher order marginals correct, so
in an adversarial setting (as is commonly employed in quantum advantage
analysis), such a result cannot be used to certify large-scale photonic
interference. Interestingly, this leads to a hierarchy of both of
verification and of spoofing, where an analysis looking at the $k-$th
order marginals can test for $k$ photon interference, but not higher.
The inadequacy of the lower-order marginals in proving large-scale
quantum interference in an adversarial setting was noted previously
by Shchesnovich \cite{Shchesnovich2020}.

\subsection{Role of photon loss}

It is interesting to note that if losses in the optical system are
identical over all modes, they commute with the interferometer, and
may therefore arbitrarily be moved to any point in the experiment.
This means that Eqn 8 is also the expression for gaussian boson sampling
with loss, since we may simply consider the marginalized photons to
have disappeared into a uniformly present loss channel. This in turn
means that the marginal distributions cannot be used to measure the
effect of photon loss onto the effective distinguishability parameter
of \cite{Renema2020ML}. 

\subsection{Arbitrary quantum states}

Kalai \cite{Kalai2014} asked whether it is possible to construct
quantum states which are resistant against the approximation method
outlined in that paper. We give an example: the quantum state $|\Psi\rangle=\frac{1}{2}\{|\psi_{1}\rangle+|\psi_{2}\rangle\},$
with $|\psi_{1}\rangle$ and $|\psi_{2}\rangle$ both of the form
of Fock state boson sampling (i.e. a product state of $n$ single
photon states and vaccum), but where the modes where these two states
have photons are completely disjunct (e.g. $|\psi_{1}\rangle$ has
photons in modes $1..n$ and $|\psi_{2}\rangle$ has them in $n+1...2n$.
In this case, the set of $\xi$ consists of two elements, $|\xi_{1}\rangle=|\psi_{1}\rangle$
and $|\xi_{2}\rangle=|\psi_{2}\rangle$, and any permutation $\sigma$
in the cross term consists of $n$-photon interference. Since there
are $n$ such terms, this means the $n$-photon interference is enhanced
exponentially compared to all others, similar to the case of random
circuit sampling.

Even more intriguingly, such a state cannot be made with a linear
interferometer, but requires many photon-photon nonlinearities (e.g.
$n$ CNOT gates, which are known to not be realizable deterministically
in linear optics). It is an interesting question whether it is the
structure of linear optics itself that imposes the noise sensitivity
noticed by Kalai, since (as the above example shows) it is possible
to remove this noise sensitivity by allowing nonlinear resources.


\begin{thebibliography}{10}

\bibitem{Zhong2020}
H.-S. Zhong {\em et~al.},
\newblock Science {\bf 370}, 1460 (2020),
  https://science.sciencemag.org/content/370/6523/1460.full.pdf.

\bibitem{Kalaicomment}
G.~Kalai,
\newblock https://www.scottaaronson.com/blog/?p=5122.

\bibitem{Boixo2017}
S.~Boixo, V.~N. Smelyanskiy, and H.~Neven,
\newblock Fourier analysis of sampling from noisy chaotic quantum circuits,
  2017, arXiv:1708.01875.

\bibitem{Kalai2014}
G.~Kalai and G.~Kindler,
\newblock Gaussian noise sensitivity and bosonsampling, 2014, arXiv:1409.3093.

\bibitem{Chaoyangcomment}
C.~Lu,
\newblock Private communication.

\bibitem{Aaronsonlostphotons}
S.~Aaronson and D.~J. Brod,
\newblock Phys. Rev. A {\bf 93}, 012335 (2016).

\bibitem{Arkhipov15}
A.~Arkhipov,
\newblock Phys. Rev. A {\bf 92}, 062326 (2015).

\bibitem{Moylett2019}
A.~E. Moylett, R.~Garc{\'{\i}}a-Patr{\'{o}}n, J.~J. Renema, and P.~S. Turner,
\newblock Quantum Science and Technology {\bf 5}, 015001 (2019).

\bibitem{Oszmaniec2018}
M.~Oszmaniec and D.~Brod,
\newblock arXiv:1801.06166 .

\bibitem{Patron2017}
R.~Garcia-Patron, J.~J. Renema, and V.~Shchesnovich,
\newblock arXiv:1712.10037 .

\bibitem{Qi2019}
H.~Qi, D.~J. Brod, N.~Quesada, and R.~Garcia-Patron,
\newblock Regimes of classical simulability for noisy gaussian boson sampling,
  2019, arXiv:1905.12075.

\bibitem{RahimiKeshari2016}
S.~Rahimi-Keshari, T.~C. Ralph, and C.~M. Caves,
\newblock Phys. Rev. X {\bf 6}, 021039 (2016).

\bibitem{Renema2018a}
J.~J. Renema {\em et~al.},
\newblock Phys. Rev. Lett. {\bf 120}, 220502 (2018).

\bibitem{Renema2018b}
J.~Renema, V.~Shchesnovich, and R.~Garcia-Patron,
\newblock arXiv:1809.01953  (2018).

\bibitem{Renema2020GBS}
J.~J. Renema,
\newblock Physical Review A {\bf 101}, 063840 (2020).

\bibitem{Shchesnovich2014}
V.~S. Shchesnovich,
\newblock Phys. Rev. A {\bf 89}, 022333 (2014).

\bibitem{Shchesnovich2015a}
V.~S. Shchesnovich,
\newblock Phys. Rev. A {\bf 91}, 063842 (2015).

\bibitem{Shchesnovich2015b}
V.~S. Shchesnovich,
\newblock Phys. Rev. A {\bf 91}, 013844 (2015).

\bibitem{Shchesnovich2017}
V.~Shchesnovich,
\newblock Partial distinguishability and photon counting probabilities in
  linear multiport devices, 2017, arXiv:1712.03191.

\bibitem{Shchesnovich2019a}
V.~S. Shchesnovich,
\newblock Physical Review A {\bf 100} (2019).

\bibitem{Shchesnovich2019b}
V.~Shchesnovich,
\newblock arXiv:1904.02013  (2019).

\bibitem{Shchesnovich2020}
V.~Shchesnovich,
\newblock Distinguishing noisy boson sampling from classical simulations, 2019,
  arXiv:1905.11458.

\bibitem{vdMeer2020}
R.~van~der Meer, J.~J. Renema, B.~Brecht, C.~Silberhorn, and P.~W.~H. Pinkse,
\newblock (2020), arXiv:2001.03596.

\bibitem{Aaronson2011}
S.~Aaronson and A.~Arkhipov,
\newblock Theory Comput. {\bf 9}, 143 (2013).

\bibitem{Clifford2017}
P.~Clifford and R.~Clifford,
\newblock The classical complexity of boson sampling, 2017, arXiv:1706.01260.

\bibitem{Walschaers2016}
M.~Walschaers {\em et~al.},
\newblock New Journal of Physics {\bf 18}, 032001 (2016).

\bibitem{Phillips2019}
D.~S. Phillips {\em et~al.},
\newblock Physical Review A {\bf 99} (2019).

\bibitem{Renema2020ML}
J.~J. Renema {\em et~al.},
\newblock Sample-efficient benchmarking of multi-photon interference on a boson
  sampler in the sparse regime, 2020, arXiv:2008.09077.

\end{thebibliography}
\end{document}